\input{aipcheck}

\documentclass[
    ,final            % use final for the camera ready runs
  ]
  {aipproc}

\layoutstyle{6x9}

\begin{document}

\title{On the Smallness of the Cosmological Constant in 
SUGRA Models Inspired by Degenerate Vacua}

\classification{04.65.+e, 95.36.+x, 12.60.Jv}
\keywords      {Supergravity, cosmological constant, supersymmetric models}

\author{C. D. Froggatt}{
  address={University of Glasgow, Glasgow, G12 8QQ, UK}
}

\author{R. Nevzorov
\footnote{Based on a talk presented by R. Nevzorov at the SUSY'09 Conference,
Boston, USA, 5-10 June, 2009}
~\footnote{On leave of absence from the Theory Department, ITEP, Moscow, Russia}~
}{
  address={University of Glasgow, Glasgow, G12 8QQ, UK}
}

\author{H. B. Nielsen}{
  address={The Niels Bohr Institute, Copenhagen, DK 2100, Denmark}
   % additional visiting address
}

\begin{abstract}
In the no--scale supergravity global symmetries protect local
supersymmetry and a zero value for the cosmological constant. 
The breakdown of these symmetries, which ensures the vanishing 
of the vacuum energy density, results in a set of degenerate
vacua with broken and unbroken supersymmetry leading to the natural 
realisation of the multiple point principle (MPP). In the MPP 
inspired SUGRA models the cosmological constant is naturally tiny.
\end{abstract}

\maketitle

%%%%%%%%%%%%%%%%%%%%%%%%%%%%%%%%%%%%%%%%%%%%
%% MAINMATTER
%%%%%%%%%%%%%%%%%%%%%%%%%%%%%%%%%%%%%%%%%%%%

\section{Introduction}

Recent observations indicate that 70\%-73\% of the energy density of the Universe 
exists in the form of dark energy. This tiny vacuum energy density (the cosmological 
constant) $\Lambda \sim 10^{-123}M_{Pl}^4 \sim 10^{-55} M_Z^4$ is responsible for 
the accelerated expansion of the Universe. In the standard model (SM) the cosmological 
constant is expected to be many orders of magnitude larger than the observed vacuum 
energy density. Indeed, much larger contributions must come from the QCD condensates
and electroweak symmetry breaking, while the contribution of zero--modes should push 
the total vacuum energy density up to $\sim M_{Pl}^4$. An exact global supersymmetry 
(SUSY) ensures zero value for the vacuum energy density. However the breakdown of
SUSY induces a huge and positive contribution to the cosmological constant of order 
$M_{S}^4$, where SUSY breaking scale $M_{S}\gg 100\,\mbox{GeV}$.

\section{MPP inspired SUGRA models}

In general the vacuum energy density in $(N=1)$ supergravity (SUGRA) models is huge and negative 
$\Lambda\sim -m_{3/2}^2 M_{Pl}^2$, where $m_{3/2}$ is a gravitino mass. The situation changes dramatically 
in no-scale supergravity where the invariance of the Lagrangian under imaginary translations and dilatations 
results in the vanishing of the vacuum energy density. Unfortunately these global symmetries also protect 
supersymmetry which has to be broken in any phenomenologically acceptable theory. The breakdown of 
dilatation invariance does not necessarily result in a non--zero vacuum energy density \cite{1}--\cite{4}. 
Let us consider a SUGRA model that involves two hidden sector fields ($T$ and $z$) and a set of 
chiral supermultiplets $\varphi_{\sigma}$ in the observable sector, which transform differently 
under the imaginary translations ($T\to T+i\beta,\,\varphi_{\sigma}\to \varphi_{\sigma},\,z\to z$) 
and dilatations ($T\to\alpha^2 T,\,z\to \alpha\, z,\,\varphi_{\sigma}\to\alpha\,\varphi_{\sigma}$).
In the considered SUGRA model the superpotential $W$ and K$\ddot{a}$hler potential $K$ can be written 
in the following form \cite{1}-\cite{3}:
\begin{equation}
\begin{array}{c}
W(z,\,\varphi_{\alpha})=\displaystyle\kappa\biggl(z^3+ \mu_0
z^2+\sum_{n=4}^{\infty}c_n z^n\biggr)+\sum_{\sigma,\beta,\gamma}\frac{1}{6}
Y_{\sigma\beta\gamma}\varphi_{\sigma}\varphi_{\beta}\varphi_{\gamma}\,,\\
K=-3\ln\biggl[T+\overline{T}-|z|^2-\sum_{\sigma}\zeta_{\sigma}|\varphi_{\sigma}|^2\biggr]+
\sum_{\sigma, \lambda}(\frac{\eta_{\sigma\lambda}}{2}\varphi_{\sigma}
\varphi_{\lambda}+h.c.)+ \sum_{\sigma}\xi_{\sigma}|\varphi_{\sigma}|^2.
\end{array}
\label{1}
\end{equation}
Here we use standard supergravity mass units: $\frac{M_{Pl}}{\sqrt{8\pi}}=1$. In Eq.~(\ref{1}) we include 
a bilinear mass term for the superfield $z$ and higher order terms $c_n z^n$ in the superpotential
that spoil dilatation invariance. We also allow the breakdown of dilatation invariance in the K$\ddot{a}$hler 
potential of the observable sector which is caused by a set of terms $\xi_{\sigma}|\varphi_{\sigma}|$
and $\eta_{\alpha\beta}\varphi_{\alpha} \varphi_{\beta}$. At the same time we do not allow the breakdown of 
dilatation invariance in the superpotential of the observable sector to avoid the appearance of potentially 
dangerous terms which lead, for instance, to the so--called $\mu$--problem and in the K$\ddot{a}$hler potential 
of the hidden sector. 

In the considered SUGRA model the scalar potential of the hidden sector is positive definite
\begin{equation}
V(T,\, z)=\frac{1}{3(T+\overline{T}-|z|^2)^2}
\biggl|\frac{\partial W(z)}{\partial z}\biggr|^2\,, 
\label{2}
\end{equation}
so that the vacuum energy density vanishes near its global minima. In the simplest case when $c_n=0$, 
$V(T,\, z)$ has two minima at $z=0$ and $z=-\frac{2\mu_0}{3}$. In the first vacuum, where $z=-\frac{2\mu_0}{3}$, 
local SUSY is broken so that the gravitino becomes massive
\begin{equation}
m_{3/2}=\biggl<\frac{W(z)}{(T+\overline{T}-|z|^2)^{3/2}}\biggr>
=\frac{4\kappa\mu_0^3}{27\biggl<\biggl(T+\overline{T}
-\frac{4\mu_0^2}{9}\biggr)^{3/2}\biggr>} 
\label{3}
\end{equation}
and all scalar particles get non--zero masses $m_{\sigma}\sim\frac{m_{3/2} \xi_{\sigma} }{\zeta_{\sigma}}$. 
In the second minimum, with $z=0$, the superpotential of the hidden sector vanishes and local SUSY remains intact, 
so that the low--energy limit of this theory is described by a pure SUSY model in flat Minkowski space. If the 
high order terms $c_n z^n$ are present in Eq.~(\ref{1}), the scalar potential of the hidden sector may have
many degenerate vacua with broken and unbroken supersymmetry in which the vacuum energy density vanishes.

Thus the considered breakdown of dilatation invariance leads to a natural realisation of the multiple point 
principle (MPP). The MPP postulates the existence of many phases with the same energy density which are 
allowed by a given theory \cite{5}-\cite{6}. In SUGRA models of the above type there is 
%When applied to $(N=1)$ supergravity, 
%MPP implies the existence of
a vacuum in which the low--energy limit of the considered theory is described by a pure supersymmetric model 
in flat Minkowski space. According to the MPP this vacuum and the physical one in which we live must be degenerate. 
Such a second vacuum is only realised if the SUGRA scalar potential has a minimum where $m_{3/2}=0$ which 
normally requires an extra fine-tuning \cite{7}. In the SUGRA model considered above the MPP conditions are 
fulfilled automatically without any extra fine-tuning at the tree--level.

\section{The value of the cosmological constant}

Because the vacuum energy density of supersymmetric states in flat Minkowski space is zero and all vacua in 
the MPP inspired SUGRA models are degenerate, the cosmological constant problem is solved to first approximation 
by our assumption. However non--perturbative effects in the observable sector can give rise to the breakdown of SUSY 
in the second vacuum (phase). If SUSY breaking takes place in the second vacuum, it is caused by the strong 
interactions. When the gauge couplings at high energies are identical in both vacua the scale $\Lambda_{SQCD}$, 
where the QCD interactions become strong in the second vacuum, is given by
\begin{equation}
\Lambda_{SQCD}=M_{S}\exp\left[{\frac{2\pi}{b_3\alpha_3^{(2)}(M_{S})}}\right]\,,\qquad
\frac{1}{\alpha^{(2)}_3(M_S)}=\frac{1}{\alpha^{(1)}_3(M_Z)}-
\frac{\tilde{b}_3}{4\pi}\ln\frac{M^2_{S}}{M_Z^2}\,, 
\label{4}
\end{equation}
where $M_S$ is the SUSY breaking scale in the physical vacuum. In Eq.(\ref{4}) $\alpha^{(1)}_3$ and $\alpha^{(2)}_3$ 
are the values of the strong gauge couplings in the physical and second minima of the SUGRA potential, while 
$\tilde{b}_3=-7$ and $b_3=-3$ are the one--loop beta functions of the SM and MSSM. At the scale $\Lambda_{SQCD}$ 
the t--quark Yukawa coupling in the MSSM is of the same order of magnitude as the strong gauge coupling. The large 
Yukawa coupling of the top quark may result in the formation of a quark condensate that breaks supersymmetry
inducing a non--zero positive value for the cosmological constant $\Lambda \simeq \Lambda_{SQCD}^4$.
The MPP philosophy then requires that the physical phase in which local supersymmetry is broken in the hidden sector 
has the same energy density as a second phase where non--perturbative supersymmetry breakdown takes place 
in the observable sector.

\begin{figure}
\begin{tabular}{c}
\hspace{-8cm}{$\log[\Lambda_{SQCD}/M_{Pl}]$}\\
\includegraphics[height=.3\textheight]{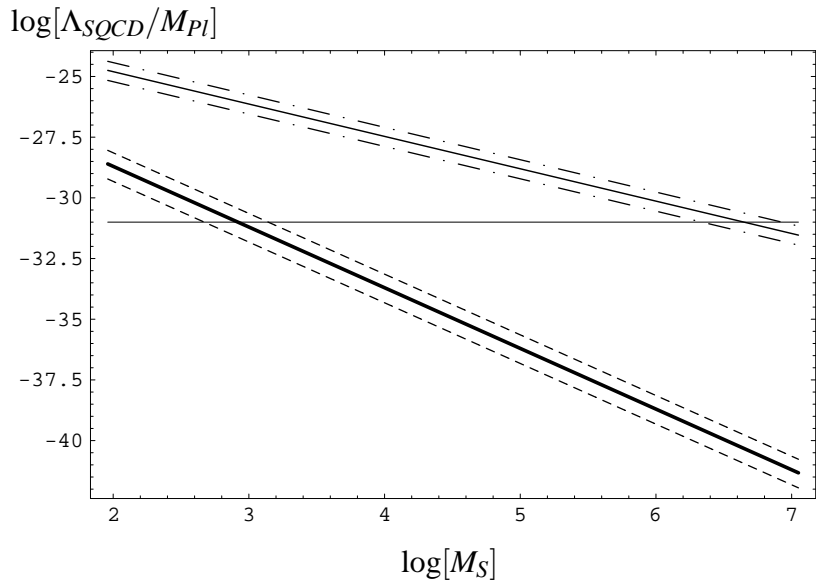}\\
\hspace{1cm}{$\log[M_S]$}
\caption{The value of $\log\left[\Lambda_{SQCD}/M_{Pl}\right]$ versus $\log M_S$.
The thin and thick solid lines correspond to the pure MSSM and the
MSSM with an extra pair of $5+\bar{5}$ multiplets. The dashed and dash--dotted
lines represent the uncertainty in $\alpha_3(M_Z)$, i.e. $\alpha_3(M_Z)=0.112-0.124$.
The horizontal line corresponds to the observed value of $\Lambda^{1/4}$. The SUSY
breaking scale $M_S$ is given in GeV.}
\end{tabular}
\end{figure}

In Fig.~1 the dependence of $\Lambda_{SQCD}$ on the SUSY breaking scale $M_S$ is examined. Because $\tilde{b}_3 < b_3$ 
the QCD gauge coupling below $M_S$ is larger in the physical minimum than in the second one. Therefore the value of 
$\Lambda_{SQCD}$ is much lower than the QCD scale in the Standard Model and diminishes with increasing $M_S$. When the 
SUSY breaking scale in our vacuum is of the order of 1 TeV, we obtain $\Lambda^4_{SQCD}=10^{-104}M_{Pl} \simeq 100$ eV
which is much smaller than an electroweak scale contribution in our vacuum $v^4 \simeq 10^{-62} M_{Pl}$. From the rough 
estimate $\Lambda \simeq \Lambda_{SQCD}^4$ of the energy density, it can be easily seen that the measured value of 
the cosmological constant is reproduced when $\Lambda_{SQCD}=10^{-31}M_{Pl} \simeq 10^{-3}$ eV \cite{1}, \cite{7}
which is attained for $M_S=10^3-10^4\,\mbox{TeV}$. However the consequent large splitting within SUSY multiplets
would spoil gauge coupling unification and reintroduce the hierarchy problem, which would make the stabilisation of the
electroweak scale rather problematic.

A model consistent with electroweak symmetry breaking and cosmological observations can be constructed, if the MSSM particle
content is supplemented by an additional pair of $5+\bar{5}$ multiplets. In the physical vacuum these extra particles would
gain masses around the supersymmetry breaking scale due to the presence of the bilinear terms $\left[\eta (5\cdot \overline{5})
+h.c.\right]$ in the K$\ddot{a}$hler potential \cite{1}. Near the second minimum of the SUGRA scalar potential the 
new particles would be massless, since $m_{3/2}=0$. Therefore they give a considerable contribution to the $\beta$
functions ($b_3=-2$), reducing $\Lambda_{SQCD}$ further. In this case the observed value of the cosmological constant can be
reproduced even for $M_S\simeq 1\,\mbox{TeV}$ (see Fig.~1) \cite{1}, \cite{7}.

\section{Conclusions}

We have argued that the breakdown of global symmetries in no-scale supergravity can lead to a set of degenerate vacua with broken 
and unbroken local supersymmetry (first and second phases) so that the MPP conditions are satisfied without any extra fine-tuning. 
In the MPP inspired SUGRA models supersymmetry in the second phase may be broken dynamically in the observable sector inducing
a tiny and positive value of the cosmological constant which can be assigned, by virtue of the MPP to all other phases.

\begin{theacknowledgments}
%RN would like to thank H.E.Haber, S.F.King, H.P.Nilles, M.Sher, D.G.Sutherland, E.E.Boos, V.A.Rubakov and I.I.Tkachev for 
%fruitful discussions. 
RN acknowledges support from the SHEFC grant HR03020 SUPA 36878.
\end{theacknowledgments}

\end{document}